# A Conversational Interface to Improve Medication Adherence: Towards AI Support in Patient's Treatment

Ahmed Fadhil
University of Trento
Trento, Italy
fadhil@fbk.eu

## ABSTRACT

Medication adherence is of utmost importance for many chronic conditions, regardless of the disease type. Engag- ing patients in self-tracking their medication is a big chal- lenge. One way to potentially reduce this burden is to use reminders to promote wellness throughout all stages of life and improve medication adherence. Chatbots have proven effectiveness in triggering users to engage in certain activ- ity, such as medication adherence. In this paper, we discuss "Roborto", a chatbot to create an engaging interactive and intelligent environment for patients and assist in positive lifestyle modification. We introduce a way for healthcare providers to track patients adherence and intervene when- ever necessary. We describe the health, technical and be- havioural approaches to the problem of medication non-adherence and propose a diagnostic and decision support tool. The proposed study will be implemented and vali- dated through a pilot experiment with users to measure the efficacy of the proposed approach.

## Categories and Subject Descriptors

H5.m. [Information interfaces and presentation (e.g., HCI)]: Miscellaneous

## Keywords

Chatbot, medication adherence, healthcare, UX design, be- haviour theory, chronic diseases

## 1. INTRODUCTION

Adherence to long-term therapy is when person's behaviour, such as taking medication, following a diet, or executing lifestyle changes, corresponds with agreed recommendations from a healthcare provider [13]. Medication adherence is a growing issue around the world. Among patients with chronic illness, approximately 50% do not take medications as prescribed. Moreover, patients who skip or forgot medica- tion intake as prescribed are at risk of health deterioration, increased hospitalisation, and mortality. The World Health Organisation WHO reports that increasing the effectiveness of adherence interventions might have a far greater impact on the health of the population than any improvement in specific medical treatment [13]. Sometimes consulting a doc- tor is also an obstacle to access the necessary advices on medication adherence.

Several approaches have been developed to tackle the is- sue of medication adherence. Research studies showed that instant text messaging could be a good solution to tackle medication non-adherence among patients [12, 8]. Unlike apps, messaging platforms represented by chatbots are sim- ple, easy to use and requires no familiarity with the tool. Moreover, since elderly people are among the population that consumes more medication and they're as well more af- fected by low medication adherence [13], using a chatbot is a perfect solution to provide them with an intuitive technol- ogy. Moreover, many elderly people are already using mes- saging applications, such as Telegram or What'sApp. Chat- bots can keep track of patient's regimen, motivate them to keep going on and can provide them with side effects in- formation. Intelligent chatbots can act as miniature doc- tors who help, through AI, to find a solution to common symptoms. However, it's worth saying that a chatbot will never live up to an experienced medical professional and can not provide the human touch to cover user emotion and be- havioural needs. The usefulness of chatbots in healthcare is their ability to provide advices and information for healthy living. For example, prompt patient to book appointment with the doctor for an actual diagnosis. Chatbots can help patients understand their symptom and can facilitate the lives of many people, making it easier to access the neces- sary health and medical information.

With the present article, we propose "Roborto", a chat- bot model system to increase medication adherence among patients and inform healthcare providers about various im- provement and deterioration in patient's condition. The bot uses natural conversation and inline options to help patients understand and manage their medication. Moreover, based on user pre-configuration, the bot sends them medication re- minder and asks them about their health condition and be- havioural state (e.g., emotion and mood). Finally, the bot design and implementation is based on a behavioural ap- proach to product design. We followed a four steps process to design Roborto chatbot. The four stages of behaviour design focus on attention, influencing decision, facilitating action and sustaining the behaviour. To the best of our knowledge, few works with scientific validations and proofs discussed chatbot application for medication adherence.

## 2. BACKGROUND

Engaging people in their health and supporting them with self-management techniques are a rational to educate patients about their health condition. However, self-management interventions often fail to show significant improvement in care. The intervention fails at different stages, due to adherence problems which can relate to user behaviour or the technology context. A work by Monkaresi et al., [9] developed an interactive environment to assist diabetes patients in positive lifestyle modification and increase their medication adherence. The study investigated different dialogue styles and interaction design aspects, informed by motivational theories and found a correlation between patients adherence and dialogue style.

Short Messaging Service (SMS) is among the widely applied approaches to increase medication adherence. SMS has been studied for a variety of disorders from medical to surgical specialities. These studies evaluated text reminder schedules ranging from multiple times a day, once daily to less frequent once weekly SMS [16]. In a review by Sarkar et al., [16] utilised SMS text to promote medication adherence for a range of health conditions including self-reported adherence, pill counts and biochemical measures. The study suggested the potential role of SMS reminders in the emerging developing economies. The findings showed that SMS reminders have been largely positive. Perhaps this enthusiasm of short message system (SMS) text messaging to engage patients in their own healthcare has been due to relatively low cost, portability and widespread use of these technologies. There is compelling evidence mHealth can improve adherence to medications and suppress viral loads in patients with chronic conditions. Lauffenburger et al., [6] conducted randomised controlled trial on patients with high blood pressure in Cape Town, South Africa. The study tested 2 different text messaging systems, unidirectional information-only messages and 2-way interactive messages, among 1372 patients of mostly black African or mixed ancestry. The study developed a bank of text-messages that they mapped to behavioural change techniques, including goals and planning, repetition and substitution, social support, and natural consequences. Chat and messaging platform offer great advantages in the context of mHealth paradigm. The rapid evolution of such technologies have opened a niche for doctor-patient (DP) communication, by providing a high population penetration rate with perfect personalisation capabilities. Abashev et al., [1] provided a model for a chatbot system that integrates machine conversation systems supplemented by natural spoken language in mHealth devices. The study suggested applying these technologies together with module apps is a good solution for a variety of translational and outpatients medicine tasks. The study concluded that using chatbot systems in doctor-patient communication can reduce costs and time on routine operations. Conversely, a study by Liu et al., [8] applied mobile text messaging and medication monitors to improve the adherence to tuberculosis (TB) treatment. The findings showed that, in China, the electronic medication monitor box improved patients adherence to take their antituberculosis drugs. However, text messaging alone, which has been shown to improve adherence to antiviral therapy among HIV-patients, did not improve medication adherence among patients with tuberculosis. The authors linked this to the messages being too frequent or too impersonal, although this intervention did reduce patient loss to follow up. Text messaging as electronic reminders provide an opportunity to improve medication adherence. In a systematic review by Sarabi et al., [14] they investigated evidence to prove whether text-message reminders were effective in improving patients' adherence to medication. The study indicated that text-messaging interventions have improved patients' medication adherence rate (85%, 29.34). The most common study design was carried out in the developed countries.

Personalised text message reminders can support mHealth interventions also among patients following release from the hospital to support their cure process and medication adherence. Park et al., [11] evaluated text messaging as reminder for medication among patients with Coronary Heart Disease CHD after being released from hospital. The result suggested mHealth may be useful for improving medication adherence during the vulnerable time for CHD patients following discharge from the hospital. Similarly, Challenor et al., [3] found that mobile text messaging was by far the most popular method patients wished to be contacted with (62%) during any mHealth intervention.

Using chat based communication comes with challenges related to the level of complexity of the language associated with the task, integrating a dialogue system with a complex backend reasoning, intention recognition as key part of the understanding process, and finally, the challenges of mixed initiative where either user or system can control the dialogue at different times to achieved effective interaction [2]. Shang et al., [15] proposed Neural Responding Machine (NRM), a neural network-based response generator for Short-Text Conversation. The NRM takes the general encoder-decoder framework, it formalises the generation of response as a decoding process based on the latent representation of the input text. The study proved NRM to generate grammatically correct and content-wise appropriate responses to over 75% of the input text. A study by Liu et al., [7] investigated evolution metrics for end-to-end dialogue systems where supervised label, such as task completion are not available. The study found these metrics's correlated weakly with human judgments of response quality in both technical and non-technical domains. The study provided future recommendations for a better automatic evaluation metrics for dialogue systems.

Discussion: The majority of studies applied Short Messaging Service (SMS) to adhere patients to their medication dosages. Moreover, there was almost no study that adapted conversational interfaces into their mHealth intervention. This could be due to the new field of conversational user interface which is still growing and researchers haven't harnessed the full potential of this promising technology as an alternative to mobile applications. Although the positive findings associated with the SMS adaption in patient support and medication adherence, to our knowledge there are several limitations associated with SMS as an mHealth tool. Conversely, adapting chatbots can provide a major shift to support mHealth interventions in this domain. To begin, SMS is an effective reminder tool, how-ever unlike chatbots, it has no educational, motivational and support capabilities. For example, a chatbot can provide more personalised reminders and instruction and provide patient adherence data to the doctor to highlight any patient improvement or deterioration with respect to the medication and disease. Short messaging services provide impersonal reminder and instructions and could be too frequent. Whereas, chatbots could provide advanced feature integration (e.g., machine learning models), its expandable, and most importantly, its cost effective, user-age friendly (e.g., easier for elderly to use), since it has low learning curve and doesn't require extra installation/configuration. Unlike SMS, chatbots can track/report patient performance to the healthcare provider. Finally, chatbots are good approaches to support healthcare systems in developing or under developed countries (e.g., in rural African villages) to provide low cost healthcare services to the citizens living there. To the best of our knowledge, the vast majority of studies focused on SMS in their services, and non have considered chatbots to provide richer and innovative solution to support their mHealth intervention.

## 3. RESEARCH GOAL

This research is highly interdisciplinary, involving health, technology and user behaviour and emotional state. One of the common rationales for using chatbots is that they support patient engagement. With this paper, we aim to establish patient-doctor interaction environment for healthcare service delivery with a simple chatbot + web application based mHealth intervention tool.

We intend to integrate the chatbot with other healthcare application for the purpose of producing personalised recommendations and tracking patient condition overtime by the healthcare provider. Thus, with this research we tend to answer the following research questions:

- How to overcome the issue of medication non-adherence?
- What is the design approach most motivating for patients?
- What are the design features with significant impact on medication adherence (e.g., digital experience, social factor, behavioural techniques)?

With these questions we will also answer the most routine issues faced by patients with low medication adherence, including exercise, nutrition, emotional state, stress level, and sleeping pattern. The chatbot is particularly focused on forgetfulness as the primary barrier to optimal medication adherence.

## 4. APPROACH

The approach is to engage healthcare providers to experience the effective use of conversational AI to empower patients to use chatbots as an enabling tool. In this way, patients can be more responsible for their health and adhered to their medication. The chatbot system offers a potential way to help patients adhere to their treatments by introducing awareness, and guidance into their routine.

One distinguishing feature in "Roborto" chatbot is the inclusion of human backup in the process. The reason is that automated systems can't respond well to patients need, specially when they feel trapped at certain point during their treatment journey. For example, when a patient asks a hybrid system for a customer support, in order for the system to redirect the user to a real agent it asks the user a set of tedious questions to better support their inquiries. Whereas, including a real healthcare provider to track patients condition can also support them with empathy while making sure users are helped quickly.

### 4.1 Chatbot Behavioural Design Process

Dealing with patients medication adherence is strongly correlated to their behaviour, therefore we should link the targeted actor of the application, with the process and output. In this way we can guide design process and evaluate the behavioural effectiveness of the chatbot. Below we will define each of the targeted actor, process and output. We then provide the stages of behavioural design to ensure the chatbot is as engaging, persuasive, and actionable as possible [17].

Actor: Who are the envisioned users who will do something different in their lives, and accomplish the intended output?. For example, patients with chronic conditions who are required to adhere to a certain medication to avoid the deterioration of their health condition.

Process: How will the patient achieve the medication adherence and what behaviour will they undertake?. For example, patients medication data will be provided to the bot, and the patient will be triggered whenever its medication time to report their adherence and other medical, behavioural and psychological conditions. This step will be mostly covered by the behaviour design process.

Output: What is the goal we intend to achieve with the chatbot?. For example, people with type 2 diabetes will be more adhered to their insulin dosages when the chatbot is in action.

#### 4.1.1 The Stages of Behaviour Design

With this step we clarify the overall behavioural vision of the chatbot, identify the patients outcomes sought, provide a list of possible actions, understand patients and what is feasible and interesting to them, and evaluate the list of possible actions. To achieve this, we will follow a four-stages framework to guide our chatbot design and development process [10](see Figure-1).

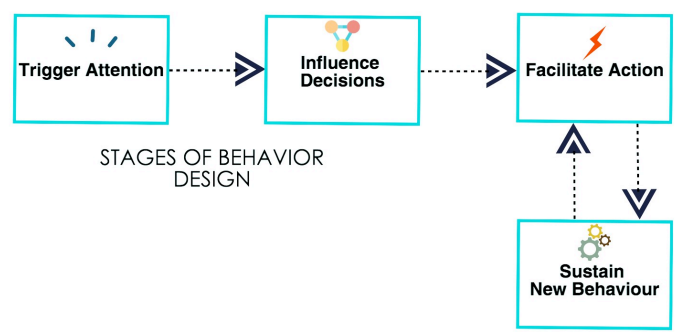

Figure 1: Chatbot Stages of Behaviour Design.

1. Trigger Attention: The first thing is to consider how to attract users to use the application. Why should

healthcare providers or patients use the chatbot?. There-fore, the bot should trigger user attention. To achieve this, we need to draw people into the bot with engaging storytelling, and compelling motion design. Moreover, the bot should provide elicit an emotional reaction, such as surprise, curiosity, or urgency. The chatbot should personalise the content and make it relevant and based on users behaviour, interest and value.

2. Influence Decisions: The chatbot has to present clear argument that nudges them to take action. Therefore, we should provide clear, straightforward content and make the message specific, simple and help patients make an informed choice. Offer recommendations and make the content compelling to patients.

3. Facilitate Action: At this stage we should help patients follow through, by creating as easy and barrier-free action as possible. This is achievable by breaking down the action into small, achievable steps. We should identify and address common barriers that prevent patients from taking action. To facilitate the action, one should guide the experience of patient in a certain way. Encourage patients to set goals and commit to actions by sending reminders and follow up on their progress.

4. Sustain New Behaviour: Finally, to achieve a long-lasting impact of medication adherence, we should motivate patients continue the behaviour and feel the sense of progress. Emphasising intrinsic motivation can be a technique to build long term behaviour change. Research shows that people are drawn to experiences that give them a sense of purpose, social connection, status, and self expression.

Applying this model of behavioural staging helps identify what stages in the journey are working and which could be better strengthening our designs and mitigating the risk of application abandonment.

### 4.2 Roborto Chatbot Functionalities

The bot functionalities need to accommodate complex data interaction, where user experience design means making a completely new concept that fits with user behaviour and condition. The chatbot supports adding a medication, where patients can identify medicine by name, icon, or photo. The bot sets a reminder about medication time and quantity. When its medication time, the bot prompts a card within the chatbot asking user to report their medication adherence, based on which they are provided with a feedback. Besides medication, the bot asks users about any symptoms and their overall stress level, emotion/mood and sleep pattern. In addition, patients can record their exercise and dietary type and other disease related information. Finally, the bot provides a report of patients condition and adherence to the healthcare providers who can track and intervene with the user via a chat window wherever relevant. The overall functionalities in Roborto bot are as summarised below:

- Evaluation of condition by asking the patient to provide a list of information to track their state of health and medication adherence.

- Tailored reminder about the necessary procedures and medication adherence, doctor appointment and visit.

- External asynchronous chat with healthcare providers through a chat channel [5].

- Doctor appointment remotely for medical procedure through a chat interface with a few button clicks.

- mHealth data collection about patient medication adherence, condition, and disease related parameters, such as sleep, feeling, diet, and physical activity. These data are forwarded to the healthcare provider who tracks their patients via a web application.

- The healthcare provider can intervene and provide patient with feedback on how to improve their overall condition, by taking into account the parameters of their behavioural patterns (nutrition, physical activity), for them to adjust their behaviour to the preset target indicators [4, 5].

- The bot can also provide information about a disease with causes and health tips and recommendations for improvements.

### 4.3 Chatbot Design

The chatbot detects the time and quantity of medication. It asks the user follow-up questions to determine their optimal reminder time. At the design phase, we focus on simple interaction and letting patients choose their optimal timing for reminder and other data entry. The bot provides quick replies to ease the bot-patient interaction. Designing a system that is smart to predict behaviour and that is utilitarian and yet engaging can get extremely hard. Beside friendliness, we also focused on motivation, empathy, discretion, and relevance while designing the bot UI. In Figure-2 we provide an overview of the chatbot medication reminder.

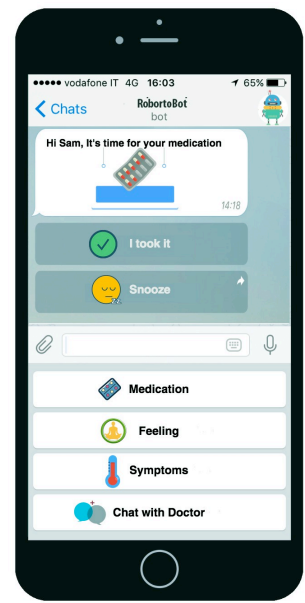

Figure 2: The Roborto Conversational Interface.

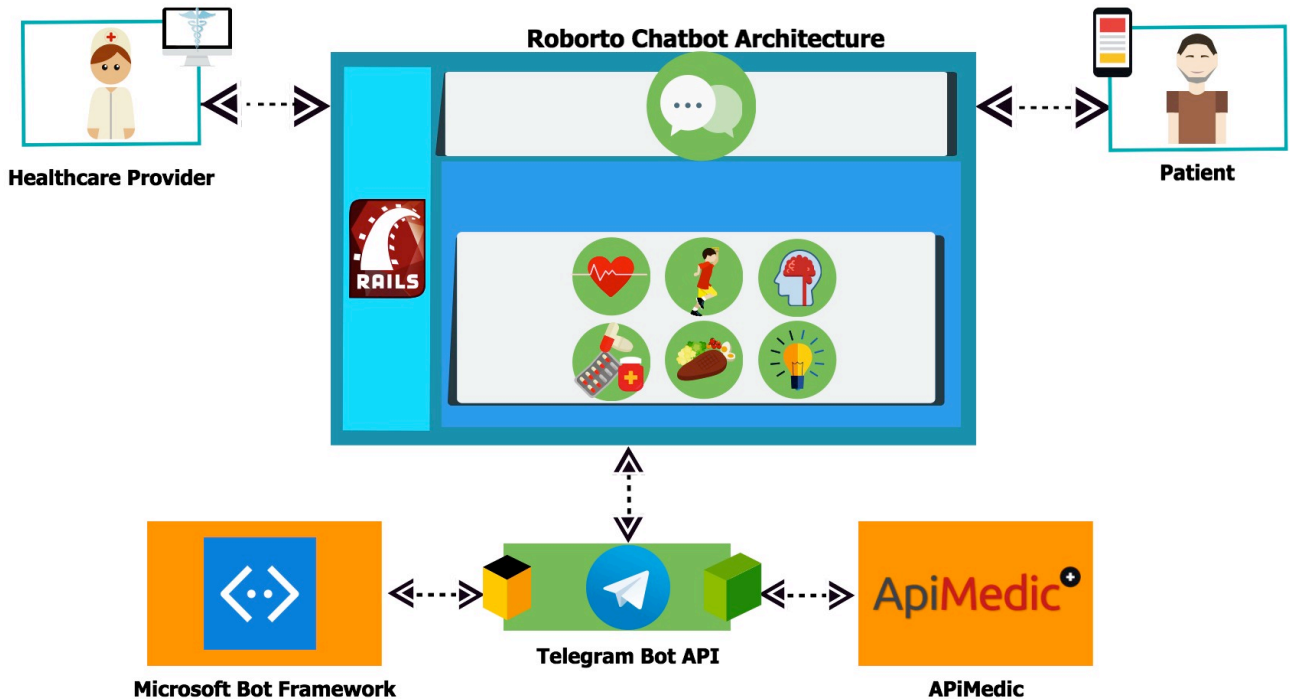

Figure 3: The Roborto Chatbot Architecture.

## 4.4 Chatbot Architecture

This phase focused on combining the technologies and de- sign aspects together to achieve efficient medication adher- ence. For the implementation, we developed a chatbot for the patient to track and report their medication, health and emotional condition, along with other lifestyle related pa- rameters (e.g., diet and exercise). However, a web applica- tion is developed for the healthcare provider to track patient condition and intervene whenever relevant.

We used the Telegram bot API[1] for the bot application to connect the bot to our system. The API provides interme- diary server that handles all encryption and communication with the Telegram API. The application can communicate with this server via a simple HTTPS-interface that offers a simplified version of the Telegram API. At the core, mes- sages, commands and requests sent by users are passed to the software running on application server. To enrich the bot in- telligence we use the Microsoft Bot Framework[2], in this way the bot acts more human like and responds more intelligently to specific user inquiries. The bot framework providers web-services with intelligence and connection using the conversa- tion channel the user authorises. The disease, symptoms and medical information were all retrieved from the APiMedic[3]. This API provides the knowledge base to our chatbot. For example, symptom checker API allows the integration of symptom checker functionalities into the bot, allowing pa- tients to find out what possible diseases they might have. The integrated application also directs users to more med- ical information. We developed the web application using Ruby and Ruby on Rails Framework[4], since it provides flex- ibility and fast development environment. In Figure-3 we provide the high-level architecture of the Roborto chatbot application architecture.

## 4.5 Contribution

There is a vast proliferation of technology available which could be useful in healthcare. In even less developed coun- tries, people are increasingly using smartphone and have ac- cess to a lot of information. However, we need a simplified way to to access these information. Conversational inter- faces offer the solution we need in this regard. A chatbot can provide patients with clinically validated information about diseases educational material or info about newly prescribed medication. All the patient has to do is start chatting with the bot.

In healthcare, a vast time is spent generating and retriev- ing information. For example, doctors diagnose based on the symptoms the patient describes, or home care services need consistent information about the health situation of their patients. A chatbot can support physicians and nurses by collecting the right information and, therefore, increase the efficiency of the overall healthcare system (quicker, cheaper, more accurate). Moreover, patients who track medication need to track other parameters, such as sleep, diet and feel- ing, a chatbot can combine all these processes and make it accessible via a chat or few button clicks.

With this research, we provide a guided conversation that provides users with predefined answers integrated into their

---

[1] https://core.telegram.org/bots
[2] https://dev.botframework.com/
[3] http://apimedic.com/
[4] http://rubyonrails.org/

Telegram chat application. Hence, we contribute by pro- viding a tool to trigger patients for medication adherence and for healthcare providers to track patients adherence and overall health condition. Moreover, rather than just a re- minder, we included a human actor into the loop to respond well to patient's emotional condition. Unlike mobile appli- cations, the chatbot has a simple interface which eliminates technology difficulties due to age barriers. Finally, since the bot collects data beyond medication adherence, it acts as a medication reminder and health condition tracker.

## 5. WHY ROBORTO CHATBOT?

The chatbot acts as a support tool and doesn't substi- tute professional medical advices. The Roborto chatbot in- tends to generate recommendations based on user informa- tion. The bot interface provides patients with reminders and follow-up questions to better predict their condition and in- form the healthcare provider about patients overall health. On the other hand, the healthcare provider is given a web application dashboard to monitor the history of patients ac- tivities, goals and medications. The aim of the developed web + chatbot application is to provide an approach and adopt the chatbot to motivate, educate and engage patients in managing their own health with a simple, cost effective technology. Hence, we believe this approach to improve ad- herence to treatment plans in patients with chronic condi- tions, through encouragement, reminders and regular chat- ting with healthcare providers. The chatbot collects data and provides instant feedback about patients condition, en- gage patients in their own health and improve outcomes, determine the cost effectiveness of this virtual clinic consul- tation strategy.

## 6. DISCUSSION

Conversational interfaces can either collect information from/for users or provide or help users with some tasks. In this paper we discussed "Roborto", a chatbot case to improve patients adherence to medication. We have a number of limi- tations that we intend to tackle in future work. For example, providing patients with information about a symptom they checked is often complicated, since there is a lot of content and its hard to scroll and read the whole text, which is often confusing. We will implement a functionality to show only part of the text and ask user to click in case they intend to read more. This way they get only as much information as they wants to read. It is crucial to consider patients feel- ing when dealing with their adherence, patients need to be supported, not blamed. For future work, we intend to in- tegrate machine learning to understand patients treatment and effective care. We also intend to integrate the bot with wearable technologies to ease the process of medication re- minder and tracking. Finally, to extract patient intention, we will use machine learning intelligence such that, when the user says a sentence as: "I am having a headache", the bot can extract user intention and provide them with a mean- ingful feedback. This could require further research, since patients often describe symptoms in very different ways and often don't know the medical terms.

## 7. CONCLUSIONS

In this paper we presented a chatbot application to adhere patients to their medication. The presented model of the chatbot system provides an innovative approach to adhere patients medication and track their condition overtime. Im- proving patients adherence might be the best approach for tackling chronic conditions effectively. A multidisciplinary approach towards adherence is needed and we need to track not only patients adherence to medication, but also their various lifestyle related activities and most importantly their emotional state. We have covered this study from technical, medical and behavioural points. The next step is to develop the complete prototype and conduct a pilot study to evalu- ate the effectiveness of the proposed system.